\documentstyle[preprint,aps]{revtex}

\begin{document}
\begin{titlepage}
\rightline{CMU-HEP94-18}
\rightline{DOE-ER/40682-72}
\leftline{hep-ph/9406}
\leftline{June, 1994}
\vspace{3cm}
\begin{center}
{\LARGE\bf SOURCES OF CP VIOLATION   \\
IN THE TWO-HIGGS DOUBLET MODEL}
\end{center}
\bigskip
\begin{center}
{\large\bf Y. L.  WU and  L. WOLFENSTEIN} \\
 Department of Physics, \\ Carnegie-Mellon University, \\
Pittsburgh, Pennsylvania 15213, U.S.A.
\end{center}
\vspace{4cm}
\begin{center}
 Phys. Rev. Lett. {\bf 73 }, 1762 (1994)
\end{center}

\end{titlepage}
\draft
\preprint{CMU-HEP94-18, DOE-ER/40682-72}
\title{Sources of CP Violation in the Two-Higgs Doublet Model}
\author{Yue-Liang  Wu\cite{byline} and Lincoln Wolfenstein}
\address{ Department of Physics, \ Carnegie Mellon University \\ Pittsburgh,
 Pennsylvania 15213,\ U.S.A.}
\date{June 1994}
\maketitle

\begin{abstract}
   Assuming CP violation arises solely through the Higgs potential, we develop
the most general two-Higgs doublet model. There is no discrete symmetry that
distinguishes the two Higgs bosons. It is assumed that an approximate global
family symmetry sufficiently suppresses flavor-changing neutral scalar
interactions. In addition to a CKM phase, neutral boson mixing, and superweak
effects, there can be significant CP violation due to charged Higgs boson
exchange. The value of $\epsilon'/\epsilon$ due to this last effect could be
as large as in the standard model.
\end{abstract}

\narrowtext

  In gauge theories the standard gauge interaction is CP invariant so that the
origin of CP violation always lies in the Higgs potential or the Yukawa
interaction of the Higgs bosons with fermions. In the standard model with only
a single Higgs doublet the only way to introduce CP violation is via
complex Yukawa couplings. The simplest extension of the standard electroweak
theory is to include two Higgs doublets instead of one. As a consequence there
exist a variety of new sources of CP violation.

  The most general Higgs potential for this case can be written

\begin{eqnarray}
V(\phi_{1}, \phi_{2}) & = & -\mu_{1}^{2} \phi_{1}^{\dagger}\phi_{1}
-\mu_{2}^{2} \phi_{2}^{\dagger}\phi_{2} -(\mu_{12}^{2} \phi_{1}^{\dagger}
\phi_{2} + h.c.) \nonumber \\
& & + \lambda_{1}(\phi_{1}^{\dagger}\phi_{1})^{2}
+ \lambda_{2}(\phi_{2}^{\dagger}\phi_{2})^{2} +
\lambda_{3}(\phi_{1}^{\dagger}\phi_{1}\phi_{2}^{\dagger}\phi_{2})
+ \lambda_{4}(\phi_{1}^{\dagger}\phi_{2})(\phi_{2}^{\dagger}\phi_{1}) \\
& &  + \frac{1}{2} [\lambda_{5}(\phi_{1}^{\dagger}\phi_{2})^{2}  + h.c. ]
+ [(\lambda_{6}\phi_{1}^{\dagger}\phi_{1}
+ \lambda_{7} \phi_{2}^{\dagger}\phi_{2})(\phi_{1}^{\dagger}\phi_{2}) +
h.c. ]  \nonumber
\end{eqnarray}
With $\lambda_{5}$ non-zero and real, CP violation can arise from non-zero
values of one or more of $\mu_{12}^{2}$, $\lambda_{6}$ or
$\lambda_{7}$. If these three (and $\lambda_{5}$) are all real, CP violation
can  occur spontaneously \cite{TDL} when $\lambda_{5} > 0$, because of
the relative phase $\delta$ between the vacuum expectation values (vevs)

\begin{equation}
<\phi_{1}^{0} > = \frac{v}{\sqrt{2}}\cos\beta  e^{i\delta}\ , \qquad
<\phi_{2}^{0} > = \frac{v}{\sqrt{2}}\sin\beta
\end{equation}
 If one of $\mu_{12}^{2}$, $\lambda_{6}$ or $\lambda_{7}$ is complex
there is explicit CP violation in the Lagrangian. In the models we
discuss in this paper we assume that the Yukawa couplings are real so that the
only source of CP violation comes from $V(\phi_{1}, \phi_{2})$. Whether the
CP violation is spontaneous or explicit the consequences of interest all depend
on the phase $\delta$ in eq.(2).

 A major issue with respect to multi-Higgs models is the possibility of
flavor-changing processes mediated by the exchange of neutral scalar
bosons (FCNE). There exist strong limits on FCNE from $K^{0}-\bar{K}^{0}$
and $B^{0}-\bar{B}^{0}$ mixing and from semi-leptonic processes like
$K_{L}\rightarrow \mu^{+}\mu^{-}$ and $B\rightarrow X \mu^{+}\mu^{-}$.
Following a theorem of Glashow and Weinberg \cite{GW} it is often proposed
to impose a discrete symmetry on the two-Higgs model under which

\begin{eqnarray}
& & \phi_{1} \rightarrow - \phi_{1}\ , \qquad \phi_{2} \rightarrow \phi_{2}\  ;
 \\
& & D_{R_{i}} \rightarrow D_{R_{i}} \qquad or \qquad D_{R_{i}} \rightarrow -
D_{R_{i}}\ ,  \qquad U_{R_{i}} \rightarrow U_{R_{i}}
\end{eqnarray}
where $D_{R_{i}}$ and $U_{R_{i}}$ are the usual right-handed quarks with
$i= 1-3$. As a result,  only $\phi_{2}$ gives mass to up quarks and
only $\phi_{1}$ or only $\phi_{2}$ gives mass to down quarks. Thus as in the
standard model the
final scalar boson couplings are proportional to the mass matrix and do not
change flavor. It also follows from  eq. (3) that the coefficients
$\mu_{12}^{2}$, $\lambda_{6}$ and $\lambda_{7}$ in eq. (1) vanish so that
no CP violation results from $V(\phi)$. Thus as in the standard model with
one doublet the only source of CP violation is the complex Yukawa couplings,
which lead to a phase in the CKM quark mixing matrix.

  Various ways of modifying the restrictions of eqs. (2) and (3) have
been proposed :

 (1)\  The discrete symmetry of eq. (3) is violated only softly by the term
proportional to $\mu_{12}^{2}$ and this is the only source of CP violation.
In order to obtain the needed CP violation in the quark sector it is necessary
to modify eq. (4) so that $d_{R_{i}} \rightarrow \eta_{i} d_{R_{i}}$ where
$\eta_{i}$ is $(+1)$ for some generations and $(-1)$ for others \cite{GEORGE}.
The consequences of such a model have been worked out in detail by
Lavoura \cite{LUIS}, he finds this is a truly superweak\cite{WOLF} model
with no CKM phase.

 (2)\  The discrete symmetry defined by eqs. (3-4) is violated both in
$V(\phi)$ and the Yukawa sector but the violation everywhere is small. This
model discussed in detail by Liu and Wolfenstein\cite{LW} also leads to
superweak CP violation but there exists in
addition a non-zero CKM phase. Furthermore, the value of $\epsilon'/\epsilon$
is greater than in generic superweak models and is expected to lie between
$10^{-4}$ and $10^{-6}$.

 (3)\  One can abandon the discrete symmetry altogether and assume that
an approximate family symmetry suppresses FCNE.
The point here is that the smallness of the off-diagonal terms in the CKM
matrix suggests that violation of flavor symmetry
(described by a set of global U(1)
transformations) are specified by small parameters. It then turns out that
reasonable choices for these small parameters combined with the natural
smallness of Higgs couplings allows one to meet the constraints on FCNE.
This point made by Cheng and Sher has recently been reemphasized
by Hall and Weinberg \cite{HW}. The consequences of this general assumption
have been worked out in
detail\cite{YLWU} by considering Approximate Global U(1) Family Symmetries
(AGUFS) (i.e., one for each family)
and is the major subject to be emphasized in this note.
Unlike Hall and Weinberg, we do not impose a particular formula for the small
parameters. Of particular importance is a new source of CP violation for
charged Higgs boson interactions that can lead to a value of
$\epsilon'/\epsilon$ as large as $10^{-3}$ independent of the CKM phase.

  After spontaneous symmetry breaking it is natural to use as a basis for the
neutral Higgs fields

\begin{eqnarray}
(v + H^{0} + iG^{0})/\sqrt{2} & = & \cos\beta \  \phi_{1}^{0} e^{-i\delta} +
\sin\beta \  \phi_{2}^{0} \nonumber \\
(R + i I)/\sqrt{2} & = & \sin\beta \  \phi_{1}^{0} e^{-i\delta} -
\cos\beta \  \phi_{2}^{0}
\end{eqnarray}
Here $H^{0}$ is the "real" Higgs boson and $G^{0}$ is the Goldstone boson
eaten up by $Z^{0}$. The orthogonal state $(R + i I)$ forms a doublet with
the charged Higgs $H^{\pm}$. The neutral mass eigenstates $H_{1}^{0}$,
$H_{2}^{0}$, $H_{3}^{0}$ are related to $(R, \  H^{0}, \  I)$ by an orthogonal
matrix $O^{H}$.

  The original Yukawa interaction has the general form

\begin{equation}
L_{Y} = \bar{\psi}_{L} (\Gamma_{1} \phi_{1} + \Gamma_{2}\phi_{2}) D_{R}
\end{equation}
plus a similar term in $U_{R}$. Here $\Gamma_{1}$, $\Gamma_{2}$ are matrices
in flavor space and $\psi_{L}$ is the quark doublet ($U_{L}$, $D_{L}$).
The assumption of Approximate Global U(1) Family Symmetries (AGUFS) says that
$\Gamma_{1}$,  $\Gamma_{2}$ have small off-diagonal elements, typically between
0.2 and 0.01 of the related diagonal element in order to fit the known CKM
matrix as well as the constraints on FCNE, i.e., AGUFS are sufficient for a
natural suppression of family-changing currents (for both charged and neutral
currents). From $L_{Y}$ one derives the mass
matrices which are diagonalized in the usual way introducing the mass basis
$u_{L}$, $u_{R}$, $d_{L}$, $d_{R}$ and the CKM matrix V.

  We now rewrite $L_{Y}$ in terms of the Higgs basis of eq. (5) and the quark
mass basis. We divide the result into a term $L_{1}$, which has no
flavor-changing effects other than that expected for $H^{\pm}$ from the
CKM matrix V, and $L_{2}$, which contains the flavor-changing effects for
neutral bosons as well as small additional flavor-changing terms for $H^{\pm}$.

\begin{equation}
L_{Y} = (L_{1} + L_{2}) \cdot (\sqrt{2}G_{F})^{1/2}
\end{equation}
with
\begin{eqnarray}
L_{1} & = & \sqrt{2} ( H^{+} \sum_{i,j}^{3}
 \xi_{d_{j}} m_{d_{j}} V_{ij} \bar{u}_{L}^{i} d^{j}_{R} - H^{-} \sum_{i,j}^{3}
\xi_{u_{j}} m_{u_{j}} V^{\dagger}_{ij} \bar{d}_{L}^{i} u^{j}_{R} ) \nonumber \\
& & + H^{0} \sum_{i}^{3} (m_{u_{i}} \bar{u}_{L}^{i} u^{i}_{R} +
m_{d_{i}} \bar{d}_{L}^{i} d^{i}_{R} ) \\
& & +  (R + i I) \sum_{i}^{3} \xi_{d_{i}} m_{d_{i}} \bar{d}_{L}^{i} d^{i}_{R} +
(R - i I) \sum_{i}^{3} \xi_{u_{i}} m_{u_{i}} \bar{u}_{L}^{i} u^{i}_{R} + H.C.
 \nonumber
\end{eqnarray}

\begin{eqnarray}
L_{2} & = & \sqrt{2} ( H^{+} \sum_{i,j'\neq j}^{3}
 V_{ij'} \mu^{d}_{j'j} \bar{u}_{L}^{i} d^{j}_{R}
- H^{-}\sum_{i,j'\neq j}^{3} V^{\dagger}_{ij'}\mu^{u}_{j'j}
\bar{d}_{L}^{i} u^{j}_{R} )  \\
& & + (R + i I) \sum_{i\neq j}^{3} \mu^{d}_{ij}
\bar{d}_{L}^{i} d^{j}_{R} +
(R - i I) \sum_{i\neq j}^{3} \mu^{u}_{ij} \bar{u}_{L}^{i} u^{j}_{R} + H.C.
 \nonumber
\end{eqnarray}
Where the factors $\xi_{d_{j}}m_{d_{j}}$ arise primarily from diagonal elements
of $\Gamma_{1}$ and $\Gamma_{2}$ whereas the factors $\mu_{jj'}^{d}$ arise from
the small off-diagonal elements.

  There are four major sources of CP violation:

  (1) \  The CKM matrix. In addition to the usual CP violation in $W^{\pm}$
exchanges there is also in all two-Higgs models a similar CP violation in
the charged-Higgs sector.

  (2) \  The phases in the factors $\xi_{f_{i}}$ provide CP violation in
the charged-Higgs exchange processes that is independent of the CKM phases.
These phases also yield CP violation in flavor-conserving $R$ and $I$
interactions.

 (3) \  The phases in the factors $\mu_{ij}^{f}$. These yield CP violation
in FCNE.

 (4)\  From the Higgs potential one derives the matrix $O^{H}$ that
diagonalizes the Higgs mass matrix. Even in the absence of fermions this
$O^{H}$ may violate CP invariance. This violation may also be described by an
invariant \cite{MP,LS} analogous to the Jarlskog invariant for the CKM matrix.
In models in which the CP violation in $L_{Y}$ is negligible this is the
major source of CP violation in effective quark interactions due to Higgs
exchange.

  A unique feature of the present analysis is the importance of the factors
$\xi_{f_{i}}$. To illustrate the origin of these factors one can simply neglect
the off-diagonal elements in $\Gamma_{1}$ and $\Gamma_{2}$ of eq. (6). (This
should be a reasonable approximation for the second and third generations
although possibly not for the first.) For example, for the third down
generation
one finds

\[ m_{3} e^{i\delta_{3}} = (g_{1}\cos\beta e^{i\delta} + g_{2}\sin\beta) v \]
where $m_{3}$ is the mass, $\delta_{3}$ is a phase associated with the mass,
and $g_{1} (g_{2})$ is the $33$ element of $\Gamma_{1} (\Gamma_{2})$.
One gets rid of $\delta_{3}$ by redefining $d_{R3}$. The corresponding coupling
of $(R + i I)$ then is derived from eqs. (5) and (6) as

\[ (g_{1}\sin\beta e^{i\delta} - g_{2}\cos\beta) v
e^{-i\delta_{3}} \equiv \xi_{d_{3}} m_{3}   \]
If $g_{1}$ and $g_{2}$ are comparable in magnitude $\xi_{d3}$ is of order
unity and has a phase like $\delta$. For example, if $\delta = \pi/2$ and
$g_{1} = g_{2}$ then the phase of $\xi_{d3}$ is $\pi/2$ independent of $\beta$.
For large values of $\tan\beta$ and $\delta = \pi/2$ one can show
$\xi_{d3} \simeq i \tan\beta \sin\delta_{3} e^{-i\delta_{3}}$ so that
for a range of $\delta_{3}$ (corresponding to a range of $g_{2}/g_{1}$)
one can obtain an enhanced value of $\xi_{d3}$ with a sizable phase.
This same factor $\xi_{d3}$ enters in the $H^{\pm}$ couplings multiplied by the
CKM matrix.

  Some of the most distinctive features of these new sources of CP violation
are

  (1)\  The factor $\xi_{f_{j}}$ provide phases in charged Higgs exchange
that can provide CP violation in tree level flavor changing amplitudes.
The important point is that these phases are in addition to and essentially
independent of the CKM phase for each particular transition. For
$\Delta S = 1$ transitions the charged Higgs boson exchange makes a
contribution to $\epsilon'/\epsilon$ which has the order of magnitude between
$10^{-4}$ and $10^{-5}$ for $\tan\beta \sim 1$ but which could be as large as
$10^{-3}$ for large
values of $\tan\beta$ (numerically, as long as $\tan\beta \sim
10 (m_{H^{+}}/200GeV)$ ) \cite{YLWU} without
conflicting with other constraints. Thus a measurement of $\epsilon'/\epsilon$
at this level would not necessarily be due to CP violation of the CKM type.

  (2)\  There may be significant contributions to $\epsilon$ from superweak
FCNE and also from box diagrams containing $H^{\pm}$.

  (3)\  The expectations for CP violation in the $B^{0}$ system can be
seriously
changed. Even if the Higgs bosons make little contributions to
$B^{0}-\bar{B}^{0}$ mixing their contribution to $\epsilon$ change the
constraints on the parameter $\eta$ \cite{WOLF2} of the CKM matrix,
allowing, for example,
the opposite sign for the $\psi K_{s}$ asymmetry\cite{SW}. It is also possible
that there may be large superweak or charged-Higgs-box-diagram contributions
to $B^{0}-\bar{B}^{0}$ mixing greatly changing the range of the asymmetries.

 (4)\  As is well-known there are many contributions to electric
dipole moments in the Higgs models of CP violation. Of
particular interest are the two-loop graphs discussed by Barr-Zee\cite{BZ}.
These contribute to the electric dipole moment $D_{n}$ of the neutron via
the chromo-electric dipole moment\cite{CG} and directly to the electron dipole
moment $D_{e}$ of the electron through the neutral Higgs boson exchanges.
In the present model because of the presence of the complex factor $\xi_{t}$
(and other $\xi_{f_{i}}$ factors), $D_{n}$ can receive a large contribution
from the Weinberg gluonic operator through the charged Higgs boson exchange and
$D_{e}$ can also receive a contribution by the same two-loop Barr-Zee
mechanism but with
virtual photon replaced by the W-boson and the neutral Higgs boson replaced
by the charged Higgs boson. The contribution to $D_{e}$ from this two-loop
diagram with charged Higgs boson exchange is comparable to that with neutral
Higgs boson exchange. From both charged and neutral Higgs boson
contributions to $D_{n}$ and $D_{e}$, values of $D_{n}$ of the order
$10^{-25}$ to $10^{-26}$ e-cm and of $D_{e}$ of the order
 $10^{-26}$ to $10^{-27}$ e-cm close to the present limits are allowed without
conflicting with other constraints.

  In conclusion, the simplest extension of the standard model, the two-Higgs
doublet model, provides rich possibilities for sources of CP violation in
addition to that from the standard CKM model. All these can arise from a single
phase between the vacuum expectation values of the two bosons. In particular,
we have emphasized the significant CP-violating effects involving exchange of
charged-Higgs bosons in a class of models in which the usual discrete symmetry
is abandoned.

This work was supported by DOE  grant \# DE-FG02-91ER40682.

\end{document}